\documentclass[twocolumn,showpacs,preprintnumbers,amsmath,amssymb,prb,superscriptaddress]{revtex4}

\usepackage{graphicx}% Include figure files
\usepackage{dcolumn}% Align table columns on decimal point
\usepackage{bm}% bold math

\begin{document}

%%\preprint{Acceptors-draft2-2007/08/21a}

\title{On the Connection of Anisotropic Conductivity to Tip Induced Space Charge Layers in Scanning Tunneling Spectroscopy of p-doped GaAs}

\author{S. Loth}
 \surname{Loth}
\author{M. Wenderoth}
 \email{wendero@ph4.physik.uni-goettingen.de}
\author{R. G. Ulbrich}
\affiliation{IV.~Physikalisches Institut der Universit\"at
G\"ottingen, Friedrich-Hund-Platz.~1, 37077 G\"ottingen, Germany}%

\author{S. Malzer}
\author{G. H. D\"ohler}
\affiliation{Max-Planck-Research Group,
Institute of Optics, Information, and Photonics,
Universit\"at Erlangen-N\"urnberg, 91058 Erlangen, Germany}

\date{\today}% It is always \today, today,

\begin{abstract}
The electronic properties of shallow acceptors in p-doped GaAs\{110\} are investigated with scanning tunneling microscopy at low temperature.
Shallow acceptors are known to exhibit distinct triangular contrasts in Scanning tunneling microscopy images for certain bias voltages. Spatially resolved I(V)-spectroscopy is performed to identify their energetic origin and behavior. A crucial parameter - the STM tip's work function - is determined experimentally. The voltage dependent potential configuration and band bending situation is derived. Ways to validate the calculations with the experiment are discussed. Differential conductivity maps reveal that the triangular contrasts are only observed with a depletion layer present under the STM tip. The tunnel process leading to the anisotropic contrasts calls for electrons to tunnel through vacuum gap and a finite region in the semiconductor.

\end{abstract}

\maketitle

\section{Introduction}
\begin{figure}
\includegraphics[scale=0.4]{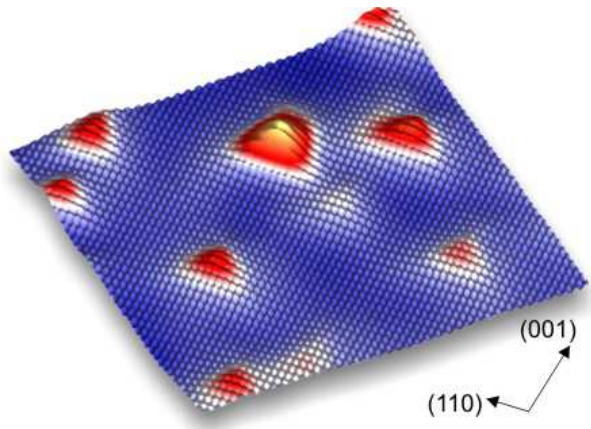}
%Man kann die Grafik auch noch beschneiden trim = 0 8 0 0,
\caption{\label{acc3d}$(40 \times40nm^{2})$ STM constant current topography of the atomically flat GaAs $(1\bar{1}0)$ cleavage surface. The corrugation of the surface states is clearly visible. At the selected sample bias +1.6V Zinc acceptors superimpose triangular shaped protrusions on the otherwise atomically flat surface. The triangles' heights vary depending on the depth of the dopant under the surface. The image shows acceptors buried up to eight monolayers.}
\end{figure}

III-V compound semiconductors have become a standard host material for fundamental research. The application of Scanning Tunneling Microscopy (STM) overcomes a  major disadvantage of far-field techniques that inherently average over a large region of the sample compared to the lattice unit cell. Non-periodic local features on the smallest length scales can be probed\cite{ros1997}. Due to the GaAs's good cleavage properties\cite{saut99} even bulk defects and buried heterostructures can be studied\cite{wie04co,win05,koe03}. Besides growing importance for future electronic devices (According to the ITRS roadmap, gate lengths will reach the 15nm range within five years\cite{itrs05}), this research field yields important information for modern solid state physics. Electronic properties of nanostructures close to the band edges can be studied  with meV-resolution via Scanning Tunneling Spectroscopy (STS)\cite{mal03,mey03}.

In the past decade the electronic contrasts induced by shallow donors and acceptors in III-V semiconductors have received extraordinary attention.  FIG.~\ref{acc3d} is a 3d visualization of a constant current topography acquired on a cleavage plane of a Zn-doped GaAs crystal. At certain sample bias voltages shallow acceptors show up as distinct triangular protrusions extending over several nanometers along the (100) directions of the crystal. The STM catches up triangles originating from acceptors that are buried up to eight monolayers under the surface. Up to now anisotropic features are found for every investigated acceptor species (reports are known for Zn\cite{zhe94,zhe94b, mah05, kor01}, Cd\cite{kor01}, Be\cite{mah05} and Mn\cite{yak04}). Additionally these features are observed in different III-V semiconductors, i.e. GaAs\cite{kor01,mah05,yak04,zhe94,lot06p}, InP\cite{kor01} and InAs\cite{mar06}. Donors on the other side  (e.~g. Si\cite{zhe94si,dom98,fee02}, Te\cite{dep99}, and Sn\cite{kor01Sn}) seem to show only circular symmetric features. They seem to comply with what is expected of an impurity hybridized with states of the nearly isotropic and parabolic conduction band.

The origin of the acceptor's triangle with the dopant site located in the triangle's tip has been subject to several reports during the past years. At first it was suggested, that d-electrons could cause such an anisotropic shape or it could be the deformation of a circular contrast due to tetragonal strain induced by the dopant atom\cite{zhe94} . Later it was argued that the triangle reflects an excited state of the acceptor wave function\cite{mah05}.

Evidently topographic measurements do not suffice to resolve this puzzle. Therefore recent research is extended to spectroscopic measurements, i.e. spatially resolved I(V)-spectroscopy carried out on p-doped GaAs. I(V)-spectra on shallow acceptors show that the conductance with triangular shape is spread out over the whole band gap interval and continues to negative voltages, as well. Recently a description has been proposed for the occurrence of this wide spread conductivity. If a depletion layer is present at the semiconductor surface, for small positive or negative bias voltages the only possible tunneling channel involves tunneling through the depletion layer. A tunneling electron is exponentially decaying in the semiconductor, i.e., it is described by the complex wave solutions of the semiconductor band gap\cite{lot06p}. For this transport channel the tunnel current is efficiently suppressed on the free GaAs surface because the depletion layer has a width of several nanometers. The acceptor's stationary negative charge locally perturbs the depletion layer and forms a double barrier potential. This results in a tunneling resonance, that enhances the tunnel current. The otherwise vanishing tunnel channel becomes visible.

Within the proposed model the presence of a depletion layer is essential for the observation of anisotropic contrasts. This paper will focus on the dependence of tip induced space charge layers at the sample's surface with the applied bias voltage between tip and sample. The chosen system is Zn-doped GaAs. We discuss the differentiation of qualitatively different tunneling situations. The results are based on the combination of measured I(V)-characteristics and numerical calculations of the band-edge distribution under the STM-tip\cite{fee02}. Energy resolved conductivity maps (dI/dV-maps) allow to attribute the occurrence of the triangular acceptor contrasts to specific tunneling conditions. In further analysis every acceptor atom within the measured sample region is identified by integrated current maps. I(V)-characteristics of the undisturbed surface are discussed within this context. The TIBB(V) calculations and the dI/dV-maps point out that the observation of triangular features at the acceptors coincides with the presence of a depletion layer at the semiconductor surface.

\section{Experiment}
The experiments are performed in a custom built low temperature STM operating in UHV at a base pressure better than $2\times10^{-11}$ mbar. The system has a lateral drift of less than ~ 0.5 \AA/h. This offers sufficient mechanical stability  for high resolution I(V)-spectroscopy. The tips are electrochemically etched from polycrystalline tungsten wire. Further processing is done in UHV by heating, sputtering and characterization with field-emission before transferring the tip into the STM. Small adsorbats that disturb the tunnel current are removed by millisecond bias voltage pulses during the measurement. This treatment results in reproducible highly stable tips which show no modifications of the apex atoms for several hours.
The GaAs samples are cleaved \emph{in situ} at room temperature and are transferred to the precooled microscope where they reach the equilibrium temperature of 8 K within less than an hour after cleavage.
The samples are conducting even at 4.2~K. The zinc doping concentration of $5\times10^{18}$cm$^{-3}$ establishes an impurity band with 10~meV spectral width centered about 30~meV above the valence band edge\cite{sch93}. At low-temperatures in bulk semiconductor, the Fermi energy is within this band and injected carriers are drained through it.

The I(V)-spectroscopies presented in this work are obtained by stabilizing the tip at a certain setpoint tunneling voltage and current and then switching off the feedback loop. The tip's position z(x,y) is recorded in the setpoint topography image. At the fixed tip sample distance z(x,y) first the apparent barrier height is measured via dI/dz-modulation technique. The tip sample distance is periodically modulated by 0.5\AA~and the modulation in the tunnel current is recorded. The apparent barrier height in eV is calculated under the assumption of exponential decay of the sample's density of States (DOS) into the vacuum and free electron mass of the tunneling particle \cite{bin84,jia98}. Secondly an I(V)-characteristic is recorded by sweeping the sample bias voltage. The exact procedure of barrier height determination and numerical processing of the raw data will be treated in detail in the next section.

\section{Numerical processing}
The measurement described above produces a data set which allows the extraction of spatially resolved apparent barrier height and energy resolved differential conductivity maps of the area of interest. At first the topographic and spectroscopic data have to be disentangled: the tip sample distance has to be fixed at a certain tunneling setpoint, the spectroscopic data is not recorded on a plane of constant height but on a corrugated surface. The height variations have impact on the I(V)-characteristics and would overlay energy dependent features\cite{gar04}. This effect is minimized in the measurement by choosing a tunneling setpoint where the topography is as smooth as possible, but it cannot be eliminated, especially when I(V)-spectroscopy is performed with atomic resolution. With the knowledge of apparent barrier height and z-position of the tip, one can numerically project the measured I(V)-curves to a plane of constant height. To increase the signal to noise ratio a gauss-weighted gliding average algorithm is applied to the I(x,y,V) data set.

After the numerics the data set is displayed in different ways. I(V)-characteristics as seen in FIG.~\ref{iuzn} give information of the overall voltage dependent behavior of certain sample regions. They are obtained by averaging over all I(V)-curves in a specified range. The red curve in FIG.~\ref{iuzn} for example corresponds to the red rectangle in the inset topography. The numerical derivative dI/dV of an I(V)-characteristic resembles the differential conductivity, i.e., the energy dependent variation of the sample's conductivity (see lower part of FIG.~\ref{iuzn}). dI/dV-maps show lateral variations in the differential conductivity at a selected sample bias. A common method to increase the signal-to-noise ratio in the band-gap region is to move the tip towards the sample during the I(V)-sweep. Usually the changing transmission coefficient of the vacuum barrier is accounted for by a normalization of the differential conductivity dI/dV with a smoothed total conductance $\overline{I/V}$\cite{fee94}. In the following sections it will become evident that the I(V)-characteristics at the acceptors exhibit non-monotonic behavior. Therefore the following measurements are carried out without tip movement and dI/dV-maps are presented to avoid artifacts created by the normalization procedure.

\section{Local I(V)-Spectroscopy}
\begin{figure}
\includegraphics[scale=0.41]{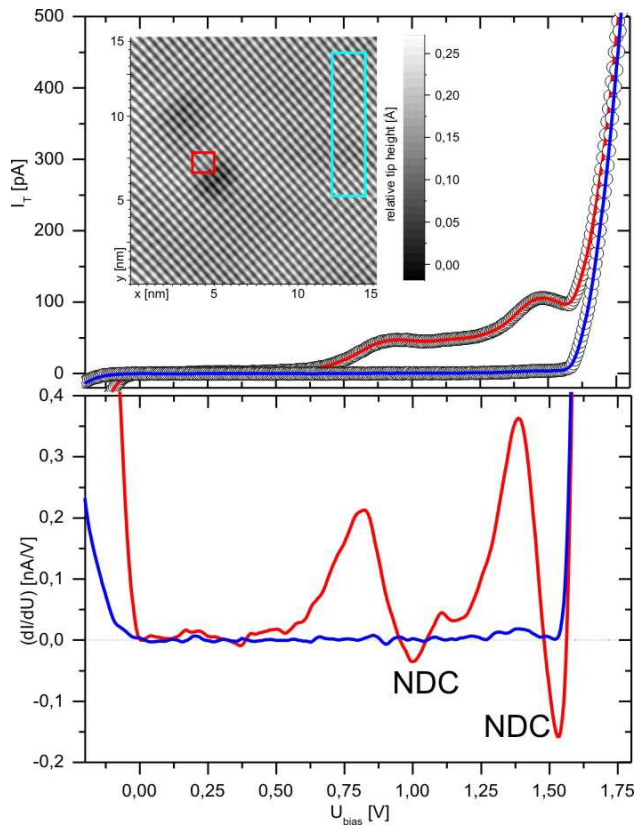}
%Man kann die Grafik auch noch beschneiden trim = 0 8 0 0,
\caption{\label{iuzn} Spatially resolved I(V)-spectroscopy. The spectroscopy is measured above a part of the clean GaAs cleavage surface. The setpoint sample bias is +1.85V and the tunnel current is set to 1nA. Two subsurface Zn acceptors are visible in the topography. They appear as faint circular depressions. No z-shift was applied. The red curve is evaluated in the small rectangle above the contrast of the lower acceptor and shows the energetic distribution of the acceptor induced conductivity. The blue curve is evaluated above the undisturbed region of the image (large rectangle). The open circles are raw data and the solid lines are the Gauss weighted average. The lower part shows the differential conductivity dI/dV of the upper two curves. Two intervals of negative differential conductivity are observed (NDC). The tunnel current above the acceptor rises up to 10\% of the setpoint current.}
\end{figure}

Figure~\ref{iuzn} presents the results of a STS-measurement performed on a zinc doped region. The upper part of the graph shows the I(V)-signal and as an inset the setpoint topography. The lower part shows the numerically derived differential conductivity. At the chosen setpoint bias voltage 1.85V and tunnel current 1nA only faint circular depressions of two buried zinc acceptors are visible. Thereby no artifacts originating from the height modulation above the acceptors occur in the I(V)-measurement. The red I(V)-characteristic is evaluated near one of these contrasts and marked by the red square in the topography. The blue curve represents an I(V)-curve evaluated above an undisturbed region of the sample marked by the blue rectangle. The data points represent raw data and the solid lines are smoothed to reduce the noise level of the dI/dV-signal as mentioned above. Both regions yield semiconductor characteristics with valence (negative voltage) and conduction (voltage larger than +1.5V) band components separated by a band gap region of about 1.5V. Above the acceptor (red curve) two groups of peaks are visible with a spectral weight of about 10\% of the setpoint current (1000pA). The tunnel current onset is as low as 0.4V. This is most prominently seen in the dI/dV signal: While the red curve exhibits drastic modulations, the blue curve on the undisturbed surface shows nearly no conductivity over the whole voltage interval from 0mV to 1550mV.
Both conductivity peaks are followed by negative differential conductivity (NDC). This points to the fact that the measurement is not only determined by the sample's local density of states. Usually the tunnel current is assumed to be the integral over the sample's LDOS weighted with an energy dependent transmission coefficient which is monotonically dependent on the applied bias voltage\cite{ham89}. Therefore the tunnel current should be a monotonic function of the sample bias. Within this approximation NDC is not explained, so a more detailed model of the tunneling process is needed.

\section{Linking Energy Scales in the Sample with the Applied Bias Voltage}

\begin{figure}
\includegraphics[trim = 20 20 20 25, scale=0.82]{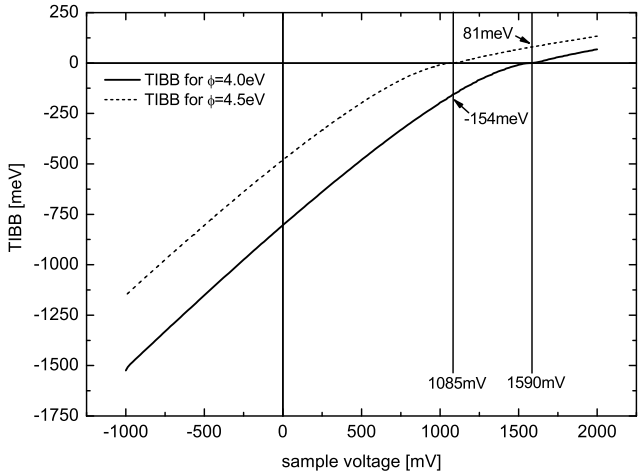}
\caption{\label{tibbcalc} Numerically derived dependence of the tip induced band bending with the applied sample bias TIBB(V). The solid curve is evaluated for the measured value of 4.0eV tip work function whereas the dotted curve corresponds to  4.5eV work function. The main difference is the sample bias where the band bending vanishes (flat band condition).}
\end{figure}

\begin{figure}
\includegraphics[scale=1.00]{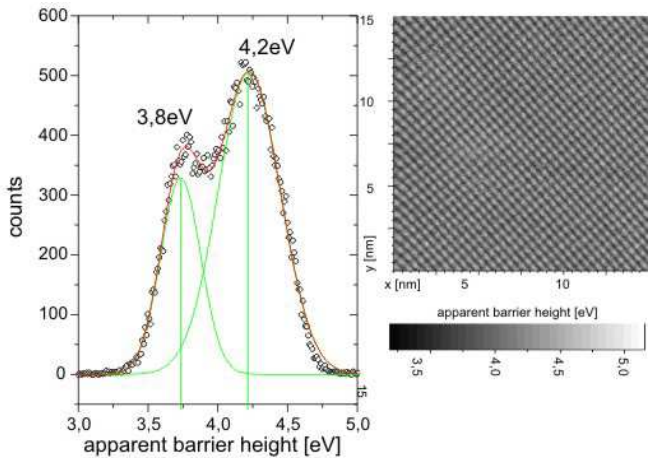}
%Man kann die Grafik auch noch beschneiden trim = 0 8 0 0,
\caption{\label{wfmeas} The tunneling barrier height is recorded for every pixel during the STS measurement via dI/dz-modulation technique. The map shows the lateral variations of the barrier height which correspond to the corrugation of the surface states. Histogramm analysis shows two peaks: 4.2eV above corrugation maxima and 3.8eV for corrugation minima. The weighted mean value is 4.05eV.}
\end{figure}

\begin{figure}
\includegraphics[trim = 20 20 0 0, scale=0.85]{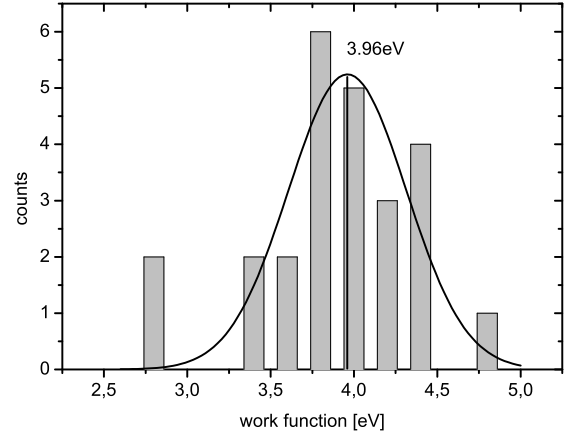}
\caption{\label{wfhist} Histogram of the mean work functions of 25 STS-measurements on the sample system. The average value for all measurements is 3.96eV work function, assuming gaussian distribution of the data points.}
\end{figure}

Above all, the understanding of such measurements with 'anomalous' behavior requires exact knowledge of the potential distribution under the tip. The unpinned Fermi energy at the GaAs\{110\} surfaces\cite{laa67} causes the electric field between tip and sample to penetrate into the crystal. This induces bias voltage dependent space charge regions under the tip. They distort the band edge alignment at the place where the measurement is performed. This effect is known as Tip Induced Band Bending (TIBB)\cite{fee87,ham93b}.

The height and sign of the surface potential (TIBB) are evaluated by solving the Poisson's equation for the situation of a metallic tip in front of a semiconductor surface. For simplicity the calculations are performed one dimensional and follow the works of Feenstra et al.\cite{fee87} and Koenraad et al.\cite{raa02}. The results are plotted in FIG.~\ref{tibbcalc}. The following parameters are used in the calculations:
The tip sample separation is approximated to 7\AA. This value is derived from previous works concerning the voltage dependent movement of the tip in z(V)-measurements\cite{lot06j} and the z-movement of the tip, when a jump to contact occurs\cite{hei97, Raa2001}.
As GaAs parameters the low temperature band gap of 1.52eV and electron affinity 4.1eV are used\cite{ada94}.
A p-doping concentration $5\times10^{18}cm^{-3}$ and 31meV ionization energy for Zn are considered\cite{sch93}. A crucial parameter for the TIBB(V) dependency is the value of the tip's work function. Figure~\ref{tibbcalc} shows two curves one evaluated for 4.0eV the other for 4.5eV work function. The tip's work function directly enters the bias value of the flat band condition, so I(V)-spectroscopies are largely affected by modifications of the tip that change the work function.

The actual value of the tip's work function for each spectroscopy can be derived from the measured apparent barrier height, if the electron affinity of the sample is known. Because the sample surface is very flat and clean it is justified to assume that the local electron affinity equals the \{110\} electron affinity of bare GaAs. Figure~\ref{wfmeas} presents the spatially resolved apparent barrier height acquired simultanously with the I(V)-spectroscopy in FIGs.~\ref{iuzn} \& \ref{didumaps}. The histogram of this barrier height map has two maxima at 3.8eV and 4.2eV respectively. They correspond to the work function values on the maxima and minima of the measured surface states. The weighted mean value is 4.05eV.
The tunnel barrier is approximated with a trapezoidal shape, like it is depicted in FIG.~\ref{didumaps}.d). Because the apparent barrier is measured by dI/dz-modulation, it is determined by the dominant transport channel at the setpoint topography. For this measurement (p-doped GaAs and +1.85eV sample bias) it is electrons tunneling from the tip into the conduction band of the sample. Electrons at $(E_F + eV)$ have the highest transmission probability\cite{ham2001}, so they determine the measured barrier height. The trapezoidal barrier is related to an effective rectangular barrier using the Wentzel Kramers Brillouin (WKB) approximation. The measured barrier height of 4.05eV is reproduced for a tip work function of 4.0eV.
It should be pointed out that the good matching of measured apparent barrier height and tip work function is only observed for the spectroscopies recorded on p-doped GaAs with setpoint sample bias below about +2V, because then the dI/dz-measurement is done nearly at flat band condition and the dominant tunneling channel is restricted to a narrow interval at the conduction band edge. When a spectroscopy is recorded at negative setpoint sample bias the measured barrier height exceeds 5eV although the tip work function is still 4.0eV. This is due to the enlarged tunnel barrier for electrons tunneling at the valence band. But still, if the band gap energy is included in the trapezoidal tunnel barrier, the tip work function can be extracted from these measurements, as well.

A tip work function value of 4.0eV work function is typical for the tungsten tips prepared by the above mentioned technique. Figure.~\ref{wfhist} shows a histogram of the mean work function values of 25 STS measurements perfomed on the p-GaAs sample system with several tips on different samples. The mean work function value of all STS-measurements is 3.96eV. It is worth noting that tip configurations with very low or very high work function values (e.g. 2.7eV or 4.7eV) show I(V)-curves with abnormal behavior. Measuring the apparent work function is a good tool to validate the actual tip configuration and ensure reproducible imaging properties.

In literature often the work function value of 4.5eV for bulk tungsten is used (see e.g. ref.\cite{mah05}). But theoretical\cite{bal02} and experimental\cite{hah80} studies reveal that this value strongly decreases when one moves from low index surfaces to higher index surfaces or increases the edge atom density on the emitting W facets. In SEM measurements our tips exhibited typical apex radii of $<10nm$. It is not surprising that the highly curved apex has a work function in the order 0.5eV lower than bulk tungsten.
To emphasize the role of the work function for the interpretation of the STS-measurements the TIBB(V) in FIG.~\ref{tibbcalc} is calculated for the measured value of $4.0eV$ (solid curve) and for $4.5eV$ (dashed curve). Both curves demonstrate that the TIBB varies monotonically with the applied bias voltage and in each bias window. For depletion (for TIBB~$<$~0meV) and accumulation (for TIBB~$>$0~eV) respectively this dependence is approximately linear. The slope of the curves depends on the charge density at the semiconductor surface that screens the tip's electric field\cite{raa02}. In the accumulation range the screening is dominated by free carriers and more efficient than in the depletion range where screening is accomplished by the immobile acceptor cores. Therefore the TIBB(V) has a steeper slope for depletion compared to accumulation. The linearity reduces a bit if the energetic structure of the GaAs DOS is increased, but the situation does not change qualitatively.

Although the overall shape of both curves is similar they differ strongly in the voltage position of flat band condition. For 4.5eV work function it is reached at 1085mV and for 4.0eV at 1590mV.
The point of flat band condition is crucial: This voltage point separates two qualitatively different tunneling regimes. At bias voltages higher than the flat band voltage the bands are bent upwards. A hole accumulation layer is formed under the tip and in the case of p-GaAs the highest states of the valence band become unoccupied. The valence band states become accessoble at positive bias. At smaller sample bias than the flat band voltage the bands are bent downwards which results in a depletion layer. For electrons with lower kinetic energy than the band gap there are no unoccupied states at the sample surface. At positive voltage the only unoccupied states left are within the acceptor band in the bulk of the crystal. The only possible tunneling channel requires that the electrons do not only tunnel through the vacuum barrier but also through the depletion layer in the GaAs crystal\cite{fee87}. Despite the change in the direction of tunneling for negative voltage the situation does not change until the sample bias overcomes the depletion layer potential.
In the spatially resolved I(V)-spectroscopy different characteristic features can be checked to validate the derived TIBB(V) and the flat band voltage position, i.e., onset of conduction band tunneling and observation of charge density oscillations.

\section{Spatially Resolved I(V)-Spectroscopy}
\begin{figure*}
\includegraphics[scale=0.85]{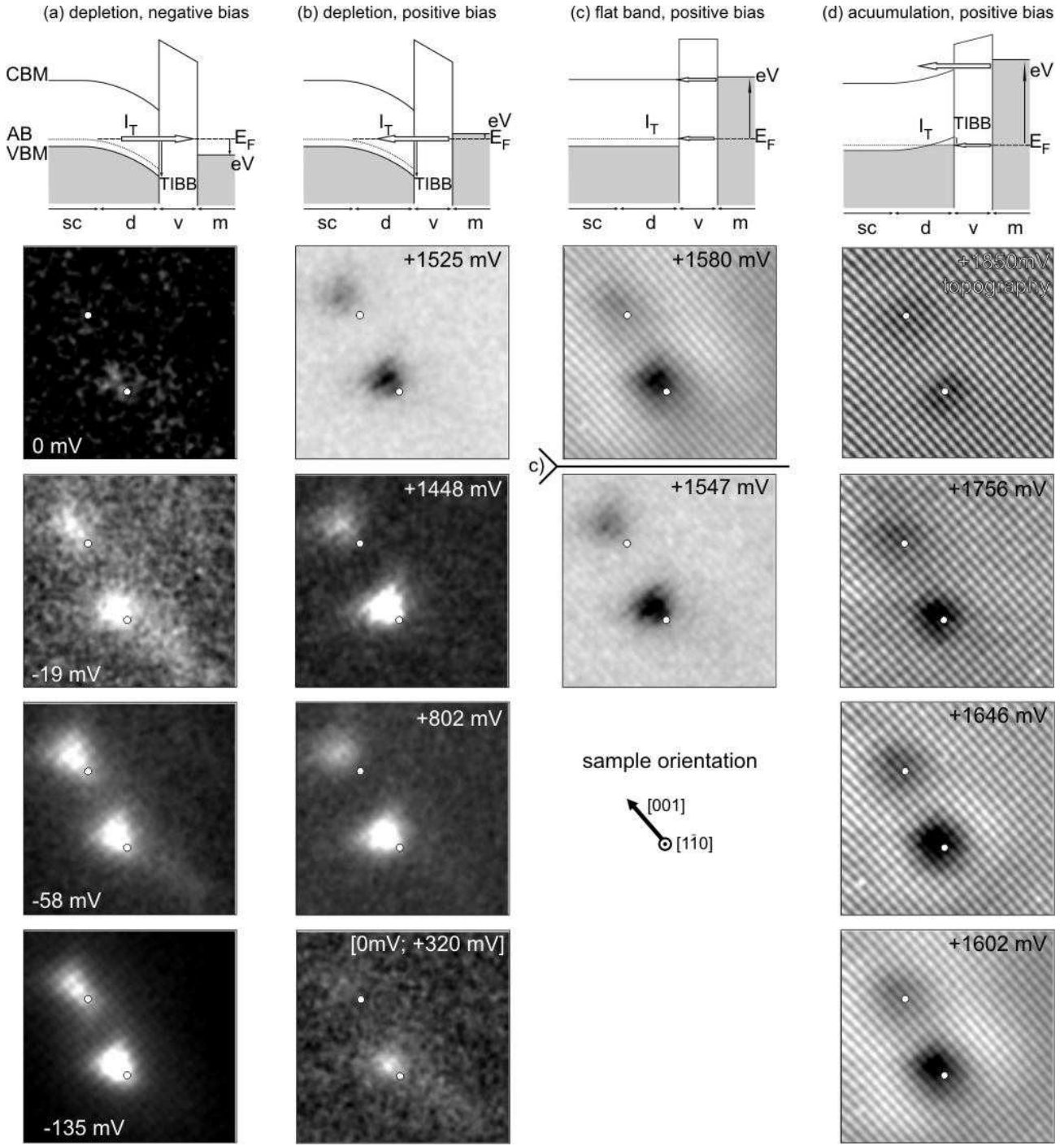}
\caption{\label{didumaps} Some representative dIdV-maps extracted from the STS-measurement shown in FIG.~\ref{iuzn}. The differential conductivity is shown in grey scale. Black is reduced and white enhanced conductivity with respect to the mean conductivity in each map. The black triangles in the maps for +1525mV to 1569mV are \textbf{negative } differential conductivity. The STS-measurement covers different tunneling situations. Each column of this figure has the same underlying potential configuration. They are sketched with rigid-band diagrams (description is given in the text): (a) depletion layer under the surface and electrons tunneling from the bulk of the sample through the depletion layer and the vacuum gap to the tip, (b) depletion layer and electrons tunneling from the tip through vacuum and depletion layer into the bulk sample, (c) flat band condition and electrons tunneling into the acceptor band and the conduction band, (d) accumulation of free holes under the surface and tunneling into the conduction band and the accumulation layer.}
\end{figure*}

The I(V)-spectroscopy presented in FIG.~\ref{iuzn} was carried out on the whole sample region of$(15 \times 15)nm^{2}$. In FIG.~\ref{didumaps} differential conductivity maps derived from this measurement are shown. Each map illustrates in grey scale the spatial variations of the differential conductivity at certain voltages. In the presented region of $(10 \times 10)nm^{2}$ two subsurface zinc acceptors are visible.  For reference the positions of the dopant atoms under the surface are indicated by white circles in each image. They are determined by the center-of-mass of the circular topographic contrast\cite{kor01}. The topography is the first image of part (d) in the figure (same topography as shown in FIG.~\ref{iuzn}). The dI/dV-maps are arranged with ascending order in bias voltage from bottom left to top right.

The computed TIBB(V) relation (FIG.~\ref{tibbcalc}) shows that four different tunneling conditions are present in the measured voltage interval. Accordingly, the dI/dV-maps are arranged in four columns FIG.~\ref{didumaps}.a), b), c) and d). The potential configuration for a tunnel path perpendicular to the surface is depicted for each interval in a rigid-band model. The schemes show the alignment of the conduction band minimum (CBM), valence band maximum (VBM) and the acceptor band (AB) with respect to the Fermi energy (E$_F$). The bulk semiconductor is denoted (sc) and the space charge region (d). The right side of each scheme illustrates the approximate shape of the vacuum barrier (v) and the tip's Fermi energy plus applied sample bias voltage (eV) in the metallic tip (m). Filled states are grey and empty states white. Possible tunneling channels (I$_T$) are selected by the condition that we assume elastic tunneling only. They are indicated by arrows. The configurations are:

(a) For negative sample bias voltages down to -135mV (FIG.\ref{didumaps}.a) the sample is in depletion. The surface potential (TIBB) is negative and the band edges in the semiconductor are pulled under the Fermi energy. Electrons are tunneling from the bulk of the semiconductor to the tip. The height of the depletion layer potential (TIBB) at the surface is larger than the applied bias voltage. At the surface no valence band states are present in the energy window of the sample DOS that can possibly contribute to the tunnel current (E$_F$ to eV in the band edge diagram). Electrons tunnel from within the sample through the vacuum barrier and also through the depletion layer in the semiconductor.

(b) For positive sample voltages up to 1590mV (calculated flat band condition TIBB(V=1590mV)=0meV, see FIG.~\ref{tibbcalc}) the depletion region persists. Despite the current direction is reversed, the situation stays the same. Until the sample bias overcomes the GaAs band gap (E$_{gap}=1.52eV$) no unoccupied states of valence or conduction band are accessible at the surface and electrons still have to tunnel through vacuum gap and depletion region (FIG.~\ref{didumaps}.b).
Shortly before the flat band condition is reached, electrons can access the conduction band ($eV>E_{gap}$) and a second tunneling channel opens.

The line (c) inicates the flat band condition at 1590mV in FIG.~\ref{didumaps}.c).

(d) For higher sample bias voltages the bands are bent upwards and the valence band maximum is pulled over the Fermi energy (FIG.~\ref{didumaps}.d). Due to the band bending free holes are accumulated under the tip. Electrons tunnel directly into the conduction band and into the valence band.

The dI/dV-maps are arranged according to this classification. The overall evolution of the measured conductivity can be divided into two intervals: In the first interval triangular features are visible at the position of the dopant atoms. They are measured for the depletion regime at positive and negative sample bias (FIG.~\ref{didumaps}.a) and b)). The triangles are positioned to the [001] side of the dopant atoms with the triangle's tip pointing to the $[00\bar{1}]$ direction. In these triangular regions the conductivity is greatly enhanced and subsequently followed by negative differential conductance as seen for the map at +1525mV. When the voltage approaches 0V from either positive or negative bias the triangles fade slowly into the background noise but do not change qualitatively. At voltages next to 0V (-19mV or up to +320mV) enhanced conductivity seems to branch away from the lower acceptor atom. This is an artifact due to another acceptor buried below the two prominent acceptors which will be discussed in part (B) of this chapter.
The second interval represents the observation of circular features at the acceptors. They appear at about +1580mV, which is at the computed flat band voltage position (FIG.~\ref{didumaps}.c)). Additionally the onset of strong surface corrugation is observed. For further increasing sample bias the circular feature becomes more prominent and shrinks towards the acceptor atoms' positions until it is no longer visible at +1756mV.

\subsection{Cross-Checking TIBB(V) with the Measurement}
\begin{figure}
\includegraphics[scale=1.00]{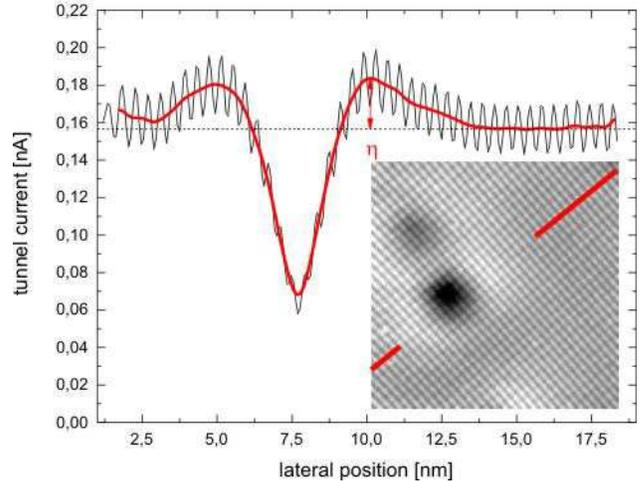}
\caption{\label{friedel} Current map derived from the dI/dV-maps representing the amount of tunnel current originating from the sample bias interval (+1547mV to +1668mV). The hole charge density oscillations are observed. The cross section directly through the center of the oscillatory pattern allows estimating that the amount of tunnel current flowing into the hole accumulation layer is about $\eta=27pA$.}
\end{figure}

The onset of the circular feature is related to an accumulation layer under the tip and allows validate the calculated TIBB(V) dependence with the measurement. When flat band condition is passed, a few meV later the valence band edge is pulled over the Fermi energy. Free holes are accumulated under the tip. They are vertically confined by the extension of the accumulation layer. The lateral confinement is determined by the dimension of the tip (for our tips 10-20nm) and much weaker. Charge density oscillations of the nearly free hole gas around charged defects are expected. Those oscillations have been previously reported for electron accumulation layers on n-type semiconductors\cite{wie96,wen99,dom99,dep99} and first indications of hole charge density oscillations (CDO) on p-GaAs have been observed\cite{kor01}. In the dI/dV-maps beginning at 1580mV a bright halo of increased conductivity surrounds the two acceptors. This halo begins with a diameter of about 10nm and decreases in size with increasing sample bias. This is the signature of a CDO in the hole accumulation layer\cite{ing00}. The TIBB increases when the sample bias is raised. The accumulation layer increases in energetic depth. More states fit into the tip induced quantum dot and the Friedel oscillation period decreases.

It is possible that even when states exist in the sample they are not imaged because tunneling into them is suppressed by low drain rates into the bulk of the sample\cite{bro00}. The question whether the accumulation layer could already be present at lower bias voltage should be addressed.
Because there is an additional tunnel channel into the conduction band for the voltage interval of interest it is worthwhile to deduce the fraction of the total tunnel current that is attributed to tunneling into the accumulation layer. A current map is derived from the I(V)-spectroscopy by integrating the differential conductivity signal over the bias voltage interval from the onset of the CDO (1547mV) to the point where they are no longer visible in the dI/dV-map (1668mv). The inset of FIG.~\ref{friedel} presents this I(V)-map. The red line indicates the position of the plotted cross section. The corrugation of the conduction band surface resonance overlays the long-range signal of the CDO. The undisturbed surface (right part of the cross section) gives a signal of 157pA. The CDO causes an additional elevation of $\eta$=27pA in the vicinity of the acceptor core. Thus the CDO has an impact of at least 17\% on the tunnel current in the respective voltage interval. This estimation shows that a significant part of the tunnel current can be driven through the accumulation layer, proving that this channel is not suppressed and once a hole accumulation layer is formed, transport into it is well observable with the STM.
The TIBB(V) calculation for 4.5eV work function would give rise to CDO already at about 1.1V sample bias. For 4.0eV the CDO is expected slightly above +1.6eV. Within the accuracy of the calculation, a measured onset of 1.58V fits well to 4.0eV tip work function.

Another way to check the surface band alignment are the onset voltages of the different surface resonances on GaAs.
According to the calculation for 4.0eV the onset of tunneling into the conduction band should begin at a bias voltage only a few meV lower than the GaAs band gap (about 1510mV) due to the negative band bending. This slight difference is not observable in the measurement with an effective resolution of about 20meV. If the tip's work function was 4.5eV the onset of tunneling into the conduction band should be observed at 1590mV. The onset of atomic corrugation with features of a conduction band surface resonance indicates this onset in the dI/dV-maps. The energetically lowest surface resonance in GaAs on the conduction band side exhibits its corrugation along a \{110\} direction\cite{che79,kor01}. With our sample alignment it is imaged as white and dark lines running from top left of the dI/dV-maps to bottom right (best seen in the topography image). The onset is observed at 1547mV, which is 43mV lower than expected for 4.5eV work function (see FIG.~\ref{didumaps.c}). This points to a work function lower than 4.5eV. But the position and width of the surface resonances is not known to sufficient accuracy. The onset of conduction band corrugation here only shows a tendency but is no proof.

Concluding this section we find that the measured value of 4.0eV apparent work function is plausible. The TIBB(V)-curve calculated for 4.0eV tip work function is in agreement with characteristic features in the measurement. The comparison of the two TIBB(V) curves demonstrates that a variation of the work function largely influences the dependence of surface potentials with applied bias, an experimental determination of this value improves the accuracy of the numerical calculations.

\subsection{Symmetry of the acceptor induced contrasts}

Now we focus on the voltage dependent features observed at the Zn acceptors in the dI/dV-measurement (FIG.~\ref{didumaps}). The topography image shows no short range distortion of the atomic corrugation. Only two faint long range depressions superimposed on the atomic corrugation are visible. This indicates that the two zinc acceptors are embedded under the two surface monolayers of the GaAs crystal. White circles indicate the positions of the Zn acceptor atoms. In the dI/dV-maps two qualitatively different features are observed: (i) The aforementioned CDO and associated circular features. (ii) Triangular features shifted along [001] to one side of the acceptor atom.

In the maps from 1580mV to 1646mV the nearly circular enhanced conductivity is visible and centered around the dopant atom's position. These feature have already been discussed in the above section and are attributed to charge density oscillations in the hole accumulation layer. For higher voltages only circular depressions remain. They are the effect of the acceptor's stationary negative charge on the host crystal's density of states. Although the acceptor core is screened by free holes in the accumulation layer, residual charge remains for distances smaller than the screening length. It is of the order of a few nanometers, here. The bands sketched for FIG.~\ref{didumaps}.d) are pushed further upwards by the negative charge. The conductivity locally decreases because the number of conduction band states available for tunneling decreases\cite{zhe94}.

For small positive bias voltage triangular regions of enhanced conductivity appear at about 350mV and are most prominent on the conductivity peaks at 802mV and 1387mV (compare to FIG.~\ref{iuzn}). The enhanced conductivity persists up to 1448mV and changes rapidly into negative differential conductivity (NDC) as seen for +1525mV and +1547mV on the same triangular region. A similar NDC is found after the first conductivity peak, as well. For negative sample bias the triangular shaped enhancement becomes clearly visible at -58mV and persists until the end of the I(V)-measurement at -135mV.
Both, enhanced and negative differential conductivity are restricted to a triangular region shifted to the [001] side of the accecptor atom with the triangle's tip pointing to the atom's position. This is true for the $(1\bar{1}0)$ and $(\bar{1}10)$ surface. When the $(110)$ or $(\bar{1}\bar{1}0)$ surfaces are imaged, this triangle is reversed and appears on the $[00\bar{1}]$ side of the acceptor\cite{mah05,lot06j}. The triangular region extends about 2nm along [001] and the triangle's base is of comparable expansion. The contrast looks similar for both polarities (tunneling into or out of the semiconductor).

The question whether the contrast symmetry does change when one passes from -58mV to +802mV is important. Previously it has been suggested that the triangular contrast resembles the squared wave function of the first excited acceptor state\cite{mah05}. This approach is based on the assumption that for very small negative voltage the ground state, which is more elongated along [001] is imaged and that the contrast changes to the triangular one for lower bias voltage. The dI/dV-map for -19mV contains enhanced conductivity in the triangular region, but for the lower acceptor a faint branch of conductivity --~hardly overcoming the noise level~-- seems to reach along $[00\bar{1}]$ away from the acceptor atom. However, this branch is displaced by about 1nm out of the symmetry axis of the triangular contrast. To get a conductivity image for very small positive voltage, the conductivity is averaged over the interval from 0V to 320mV (see lower image in FIG.~\ref{didumaps}.b)). The resulting dI/dV-map looks similar to the one for -19mV.

\begin{figure}
\includegraphics[scale=1.00]{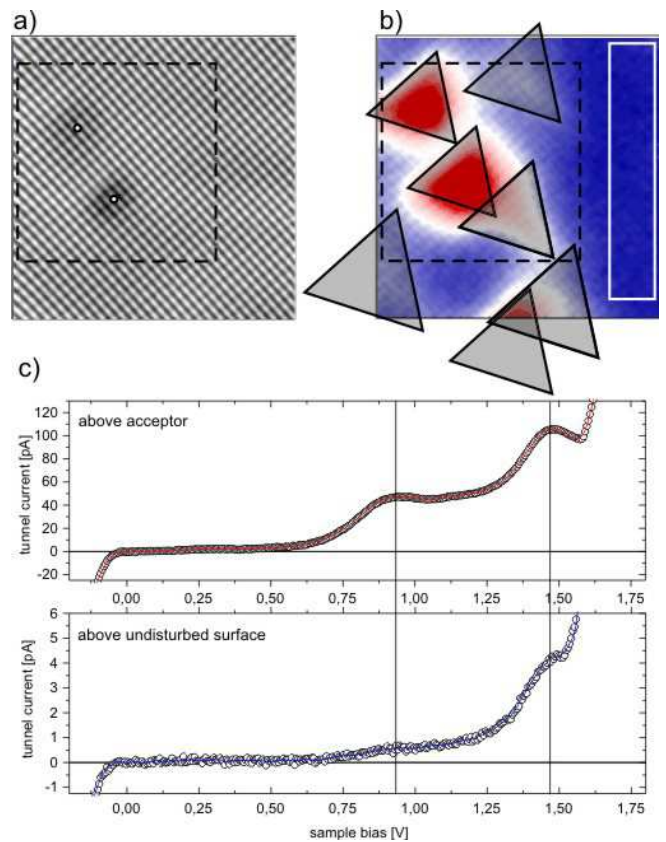}
\caption{\label{accident} Integral current map of the STS in FIG.~\ref{iuzn}. The tunnel current is integrated over the voltage interval from 0V to 1.5V and plotted color coded. The area inside the dashed rectangle in (a) is the same region as shown in the dI/dV-maps of FIG.~\ref{didumaps}. The integral current map allows the identification of all subsurface acceptors by their triangular contrast, even of those not visible in constant current topographies. The upper I(V)-curve in (c) is reproduced from FIG.~\ref{iuzn}, the lower one is an averaged curve above the undisturbed region marked in (b) with the white rectangle. The ordinate axis is magnified by a factor of $\sim$20 with respect to the upper plot.}
\end{figure}

This apparent change in contrast symmetry for the "zero bias" conductivity maps is traces back to the overlap of two acceptor contrasts. The prominent one originates from an acceptor near the surface layers and the faint one from an acceptor buried deeper under the surface. Its contrast is also present in all other dI/dV-maps presented in FIG.~\ref{didumaps} (e.g. for -58mV) but not always recognizable due to the images' color scale. An integral current map showing the tunnel current integrated from 0V to 1.5V is used to identify all subsurface acceptors that influence the measurement. This map is shown in FIG.~\ref{accident}.b). For comparison the topography is plotted, too (FIG.~\ref{accident}.a). The sample area shown in FIG.~\ref{didumaps} is marked with a dashed square. In the integrated current map even those acceptors are resolved that could not be identified in constant current topographies. As a guide-to-the-eye the identified acceptor contrasts are overlayed with grey triangles. The two acceptors nearest to the surface influence the setpoint topography. They are visible as faint depressions. The integral current map resolves seven acceptors in the region of the setpoint topography. Four of them are within the area of FIG.~\ref{didumaps}.

According to the presented I(V) measurements there is no change in contrast symmetry observable for either small positive or negative voltage. In the depletion interval from -0.1 to +1.5V only triangular contrasts are observed near the acceptor atoms. In addition the direction of tunnel current has no impact on these features. They persist over a voltage range of about 2V rather than being located at certain voltages. It is unlikely that the triangular contrast is solely attributed to the imaging of a specific acceptor state's wave function, at least for the presented I(V)-spectroscopy. This is strong indication that additional tunneling processes are involved.

\subsection{I(V)-characteristics of the undisturbed surface}
The influence of deeply buried acceptor contrasts on the dI/dV-maps makes it interesting to examine the part of the sample where no separate acceptor contrast can be identified (white rectangle in FIG.~\ref{iuzn}). The average I(V)-characteristic of this region is plotted in the lower graph of FIG.~\ref{accident}.c). The I(V)-curve recorded above the acceptor resonance is plotted in the upper part for reference. In the band gap interval from 0V to 1.5V there is faint tunnel current detected $<5pA$ even on the undisturbed surface. Comparison with the measured acceptor induced conductivity reveals that the I(V)-curve possesses the same characteristic shape. We conclude that this is due to acceptors buried deep below the surface. No separate contrasts are visible, but rather a homogenous background of acceptor induced conductivity is detected. It is present on the whole surface, so the tunnel current is influenced by an acceptor at every point of the investigated sample. This is consistent with the high doping density of the sample.

\subsection{Summary}
Bringing together the TIBB(V) calculations and the dI/dV-maps we find that the observation of triangular features at the acceptors coincides with the presence of a depletion layer at the semiconductor surface. The characteristic shape of the acceptor induced tunnel current is also present at the undisturbed surface, which indicates that this transport mechanism is active on every point of the sample. For bias windows when the anisotropic features are observed, the only possible tunneling channel requires tunneling not only through the vacuum gap but also through this depletion layer. Neither conduction band states nor valence band states are accessible directly at the surface. These observations link the triangular contrasts in the band gap window with a transport channel involving tunneling through a finite region in the semiconductor.
When unoccupied sample states (bulk valence or conduction band or surface resonances) become accessible for tunneling at the surface, the triangular features disappear. This is measured as negative differential conductivity.  The band related tunnel current shows only circular symmetric distortions at the acceptors, that are caused by the acceptor core's negative charge.

\section{Conclusion}
Spatially resolved I(V)-spectroscopy has been employed to study the electronic properties of shallow acceptors in p-doped GaAs. Against the background of several reports about the distinct triangular acceptor contrasts we focussed on the determination of the conditions in the sample that lead to these contrasts. Differential conductivity maps of two acceptors are analyzed with respect to the potential configuration under the STM tip. The triangular shape conductivity is spread over a large bias window of about 2V width. The band alignment for this bias range is derived on the basis of TIBB calculations. It is the key finding that the sample is in depletion when the triangular shaped conductivity is observed.
The tip's work function proves to be a crucial parameter for the TIBB calculation and in turn for the analysis of the I(V)-spectroscopies. The experimentally determined 4.0eV work function for the tip apex is significantly lower than the often used 4.5eV for bulk tungsten.

\begin{acknowledgments}
This work has been supported by the DFG, SFB 602 - \emph{Complex structures in condensed matter from atomic to mesoscopic scales}, and the German National Academic Foundation.
\end{acknowledgments}
%\bibliography{d:/STMworks/lothi}

\end{document}